\documentclass[pra,twocolumn,showpacs,preprintnumbers,superscriptaddress]{revtex4}
\usepackage{amsmath}
\usepackage{amssymb}
\usepackage{graphicx}
\usepackage{epstopdf}
\usepackage{textcomp}
\usepackage{subfigure}
\usepackage{hyperref}
\usepackage{enumitem}
\usepackage{etoolbox}
 \usepackage{mathrsfs}
\usepackage{color}

\DeclareMathOperator{\tr}{Tr}

\begin{document}

\title{Complementarity in Quantum Information Processing Tasks}
\author{Jaya Chaubey}
\affiliation{Center for Computational Natural Sciences and Bioinformatics,} 
\author{Sourav Chatterjee}
\affiliation{Center for Computational Natural Sciences and Bioinformatics,}
\author{Indranil Chakrabarty}
\affiliation{Center for Security, Theory and Algorithmic Research, International Institute of Information 
Technology-Hyderabad, Gachibowli, Telangana-500032, India.}

\begin{abstract}

Complementarity have been an intriguing feature of physical systems for a long time. In this work we establish a new kind of 
complimentary relations in the frame work of quantum information processing tasks. In broadcasting of entanglement we create 
many pairs of less entangled states from a given entangled state both by local and non local cloning operations. These entangled 
states can be used in various information processing tasks like teleportation and superdense coding. Since these states are 
less entangled states it is quite intuitive that these states are not going to be as powerful resource as the initial states. 
In this work we study the usefulness of these states in tasks like teleportation and super dense coding. More precisely, 
we found out bounds of their capabilities in terms of several complimentary relations involving fidelity of broadcasting. 
In principle we have considered general mixed as a resource also separately providing different examples like a) Werner like 
states, b) Bell diagonal states.  Here we have used  both local 
and non local cloning operations as means of broadcasting. In the later part of our work, we extend this result 
by obtaining bounds in form of complimentary relations in a situation where we have used $1-N$ cloning transformations instead of 
$1-2$ cloning transformations. 

\end{abstract}

\maketitle

\section{Introduction}
\noindent For the last two decades we have seen that information processing with quantum systems gives us a huge advantage 
from its classical counterparts. Quantum entanglement \cite{einstein,horodecki-entanglement} which lies at the heart of
quantum information theory is one of the key reason for such technological leap.
It plays a significant role in computational and communicational processes like quantum key
generation \cite{ekert,ben1,ben4}, secret sharing \cite{secretSharing}, teleportation \cite{ben2,pati,horo,chak,vers}, 
superdense coding \cite{ben3}, entanglement swapping\cite{bose}, remote entanglement distribution \cite{sazim-tele}, 
broadcasting \cite{buzek3, buzek2,kar,adhikari} and in many more tasks \cite{entanglement-others}. In one hand quantum information theory allows us to do so many things that can not be done 
with classical information processing systems while at the same time it also forbids us from doing several operations 
which are otherwise possible with classical systems. One such fundamental impossibility in this context is the inability 
to clone quantum states \cite{wootters}. This can be put in form of a  theorem called ``No cloning theorem" which states that there does not exist any unitary operation that will take two distinct non-orthogonal quantum states 
($|\psi \rangle$, $|\phi\rangle$) into states $|\psi\rangle\otimes|\psi\rangle$, $|\phi\rangle\otimes|\phi\rangle$ 
respectively.
Even though we cannot copy an unknown quantum state perfectly but quantum mechanics never rules out the 
possibility of cloning the state at least in an approximate manner \cite{wootters, buzek1, brub, buzek3, cloning-others, gisin, gisin2,cerf1, cerf2, duan, clone-review}. 
Probabilistic cloning is also another possibility where one can always clone an arbitrary quantum state perfectly with some 
non vanishing probability of success \cite{duan, clone-review}.\\

\noindent The term broadcasting can be used in different perspectives like broadcasting of states, broadcasting
of entanglement and more recently in broadcasting of quantum correlation. 
 Barnum et al. were the first to talk about the broadcasting of states where they showed that 
non-commuting mixed states do not meet the criteria of broadcasting \cite{barnum-noncommu}. 
Many authors showed by using sophisticated methods that 
correlations in a single bipartite state can be locally broadcast if and only if the states are classical in nature 
(i.e. having classical correlation) \cite{piani, barnum-gen, luo, luo-li}.  When we refer broadcasting of an 
entangled state, we mean creating more pairs of less entangled state from a given entangled state. One way of doing this is 
by applying local cloning transformations on each qubit of the given entangled state \cite{buzek3, buzek2}. This can also be done by 
applying global cloning operations on the entangled state itself \cite{kar}.In their paper Buzek et al. showed that 
indeed the decompression of initial quantum entanglement is possible by doing local cloning operation \cite{buzek2}. 
Further, in a separate work Bandyopadhyay et al. \cite{kar} showed that universal quantum cloners 
with fidelity is greater 
than $\frac{1}{2}(1+\sqrt{\frac{1}{3}})$ are only suitable as the non-local output states becomes 
inseparable for some values of the input parameter $\alpha$. In addition to this they proved that an 
entanglement is optimally broadcast only when optimal quantum cloners are used. They also showed that broadcasting of 
entanglement into more than two entangled pairs is not possible using only local operations. Ghiu investigated the 
broadcasting 
of entanglement by using local $1\rightarrow2$ optimal universal asymmetric Pauli machines and showed that the 
inseparability is optimally broadcast when symmetric cloners are applied \cite{ghiu}.  In other works, authors 
investigated the problem of secretly broadcasting of three-qubit entangled state between two distant partners with universal quantum cloning machine and then the result is generalized to generate secret entanglement among 
three parties \cite{satya-broad2}. Various other works on broadcasting of entanglement depending on the types of 
QCMs were also done in the later period \cite{indranil-broad, satya-continuous-broad}. In another recent work we studied 
whether we can broadcast quantum correlation that goes beyond entanglement. In a recent work authors have investigated 
the problem of broadcasting of quantum correlation \cite{discord,modi-review,dissension,zhang, discord-luo, dakic, girolami,entanglement-geometric} 
that goes beyond the notion for general two qubit states \cite{sourav}.\\ 

\noindent The problem of complementarity or mutually exclusiveness of quantum phenomenons was from the beginning of
quantum mechanics. Following the discovery of Heisenberg uncertainty principle \cite{Heisen} it was Bohr who came up with the concept of 
complementarity in the following year \cite{Bohr}. Even in quantum information theory mutually exclusive aspects of physical phenomenon 
is not something new as there had been previous instances depicting the complementarity between the local and non local information of 
the quantum systems \cite{Oppen} and between the correlation generated in dual physical processes like cloning and deletion \cite{sazim-cloning}.

\noindent We do broadcasting when we require more number of entangled states in a network instead of a highly entangled state. This is mostly required when we require to do distributed information processing tasks. It is natural to expect in a bipartite situation that these newly born entangled states are less suitable in tasks like teleportation and super dense coding than the parent state. However it is not well known how their capability are controlled by our ability of broadcasting. In this work we find out several complimentary relations manifesting the interdependence of their information processing capabilities and fidelity of broadcasting. We extend our investigation in a situation where instead of using $1-2$ cloning, we have used $1-N$ cloning transformations both locally and non locally. Eventually we find out the fidelity of broadcasting and the change in the information processing capacities for different values of $N$ ($N=3,4,5$) and investigate how these complimentary relations behave with the increase in number of copies ($N$). 

\noindent In section 2 we briefly describe the standard procedure of broadcasting of quantum entanglement with the aid of both local and non local cloning machine. In section 3 we present the complimentary relationship between the fidelity of broadcasting and the capability of information processing tasks like teleportation and super dense coding. First of all we provide numerical results in form of various plots for general two qubit mixed states. Then we have considered particular examples like Werner like states and Bell diagonal state and show the complimentary phenomenon in each of these examples. In section 4 we separately study complimentary phenomenon with increased number of obtained copies by applying general optimal $1-N$ cloning transformation instead of $1-2$ cloning transforms. 

\section{Optimal 1 to many quantum cloning machines}
\label{sec:cloningmachines}

\noindent Quantum cloning transformations can be viewed as a completely positive trace preserving map between two quantum systems, supported by an ancilla \cite{brub, clone-review}. In this section, firstly we revisit the Gisin-Massar (G-M) QCM which we will later use for our complementarity analysis with the entangled output states in the broadcasting process via local cloning \cite{gisin}. More particularly, it is an optimal state-independent QCM which creates $N$ identical copies from an input qubit. When $N=2$, the G-M cloner reduces to the Buzek-Hillery (B-H) local state-independent optimal cloner \cite{buzek1, gisin}. Secondly, we extend the idea of G-M cloner to a state independent two dimensional nonlocal cloner like the B-H QCM in higher dimensions which copies a general two qubit input state into $N$ identical copies with an optimal fidelity \cite{gisin, buzek3}.\\

\noindent \textbf{Local state-independent $1 \rightarrow N$ cloning \cite{gisin}:} The unitary operator $U_{1,N}$ QCM is described as:
\begin{eqnarray}
U_{1,N}|\uparrow>\otimes M =
\sum_{j=0}^{N-1}\alpha_j|(N-j){}\uparrow,j{}\downarrow>\otimes M_j
\,\,\,\,\,\,\,\,\,\,\,\,\,\,\,\,\,\,
\nonumber\\
U_{1,N}|\downarrow>\otimes M = \sum_{j=0}^{M\!-\!1}
\alpha_{N\!-\!1\!-\!j}|(N\!\!-\!\!1\!\!-\!\!j)\uparrow,(j\!+\!1)
\downarrow>\otimes M_j
\label{eq:Udown}
\end{eqnarray}
where $\alpha_j=\sqrt{\frac{2(N-j)}{N(N+1)}}$, $M$ denotes the initial state of the copy machine, $N-1$ denotes the blank copies, $M_j$ represent the ortho-normalized internal states of the QCM. Here, the symmetric and normalized states are given by $|N-j\psi, j \psi^{\perp}>$ with $N-j$ qubits in the state $\psi$ and $j$ qubits in the orthogonal state $\psi^{\perp}$. 

\noindent A combinatorial series calculation illustrates that
this unitary operator acts on an arbitrary input state $\psi$ as:
\begin{eqnarray}
U_{1,N}|\psi>\otimes M =
\sum_{j=0}^{N-1}\alpha_j|(N-j){}\psi,j{}\psi^\perp>\otimes M_j(\psi),
\label{eq:localcloner}
\end{eqnarray}
where $M_j(\psi)$ is the internal state of our QCM with $M_j(\psi)\perp M_k(\psi)$ for all $j\ne k$. With $\psi^* = cos \theta /2 |\uparrow^*> + e^{-i \phi} sin \theta/2 |\downarrow^*>$ which transforms under rotation as a complex conjugate representation, we get a general expression for $M_j(\psi)$. When we recognize the machine states of the QCM $M_j$ with the states $M_j = |(N-1-j){}\uparrow^*,j{}\downarrow^{*}>$, then the states $M_j(\psi)$ become $M_j(\psi)=|(N-1-j){}\psi^*,j{}(\psi^*)^\perp>$.\\

\noindent \textbf{Nonlocal state-independent $1 \rightarrow N$ cloning :}  The nonlocal version ($U_{1,N}$) of the above QCM is described as,
\begin{eqnarray}
&& U_{1,N} \left| \psi_i \right> \otimes \Sigma^{ (N-1)} \otimes M = \nonumber\\
&& \sum\limits_{j=0}^{N-1} \alpha_j  \sum\limits_{k = 0}^{n} \left| (N-j) \left|\psi_i \right> , j \left| \psi_l \right> \right> \otimes M_{jl},
\label{eq:nonlocalcloner}
\end{eqnarray}

\noindent where with $i = \left\{1,2,3,4\right\}$ we get $\psi_1 = 00,\, \psi_2 = 01,\, \psi_3 = 10,\, \psi_4 =11$; and $\Sigma^{\otimes (N-1)}$ represent $N-1$ blank states. Moreover, $M$ is the initial machine state, $M_{jl}$ are the machine state after cloning, $\alpha_{j}$ is the probability that there are $j$ errors out of $N$ cloned states.
$\left| (N-j) \left|\psi_i \right> , j \left| \psi_l \right> \right>$ represents a normalized state in which $(N-j)$ states are in $\psi_i $ state and $j$ states are in $\psi_l$ state $s.t.\ l \neq i$. Similarly, the machine states 
after cloning can be represented as
$M_{jl} = \left| (N-1-j) \left| \psi_i \right> , j \left| \psi_l \right> \right>$
where, $n$ = No. of ways in which $j$ states can be chosen out of available basis set $\psi_i$, where $i = 1,2,3,4$ such that $i \neq l$.
$\alpha_j = \frac{(N-j)*n}{\sum\limits_{j=0}^{N-1}(N-j)*n}$.
  
\section{Broadcasting of Quantum Entanglement}
\label{sec:broad_entanglement}

\noindent In this section, we briefly discuss the principle of broadcasting of quantum entanglement (inseparability) with the help of both local and nonlocal cloning operations. Let us start with a two qubit mixed state $\rho_{12}$ shared by two distant parties $A$ and $B$  which can be canonically expressed as \cite{gisin2}:
\begin{eqnarray}
\rho_{12}&=&\frac{1}{4}[\mathbb{I}_4+\sum_{i=1}^{3}(x_{i}\sigma_{i}\otimes \mathbb{I}_2+ y_{i}\mathbb{I}_2\otimes\sigma_{i})\nonumber\\
&+&\sum_{i,j=1}^{3}t_{ij}\sigma_{i}\otimes\sigma_{j}]=\left\{\vec{x},\:\vec{y},\: T\right\}\:\:\: \mbox{(say),} \label{eq:mix}
\end{eqnarray}
where $\overrightarrow{X}=\left\{ x_{1},\: x_{2},\: x_{3}\right\}$ and $\overrightarrow{Y}=\left\{ y_{1},\: y_{2},\: y_{3}\right\}$ are Bloch vectors with $0\leq\left\Vert \overrightarrow{x}\right\Vert \leq1$ and $0\leq\left\Vert \overrightarrow{y}\right\Vert \leq1$. Here, $t_{ij}$'s ($i,\:j$ = $\{1,2,3\}$) 
are elements of the correlation matrix ($T=[t_{ij}]_{3 \times 3}$) and  $\sigma_i=(\sigma_1,\sigma_2,\sigma_3)$ are the Pauli matrices with $I$ being the identity matrix.\\

\noindent The basic motivation is to broadcast (decompress) the amount of entanglement present in the given input pair to more pairs. First, we apply local (given by Eq.~\eqref{eq:localcloner}) or nonlocal (given by Eq.~\eqref{eq:nonlocalcloner}) cloning operations on the two qubit state $\rho_{12}$, given in Eq.~\eqref{eq:mix}, to produce a composite system $\tilde{\rho}_{1234}$. The broadcasting of quantum entanglement will be possible if we are able to produce more distant entangled pairs without producing local entangled pairs. In other words, if local outputs states  $\tilde{\rho}_{13}$ and $\tilde{\rho}_{24}$ obtained after cloning are separable while nonlocal output states $\tilde{\rho}_{14}$ and $\tilde{\rho}_{23}$ are inseparable, then we will be able to claim that we have created more entangled pairs $\tilde{\rho}_{14}$, $\tilde{\rho}_{23}$ from the initial pair $\rho_{12}$.\\
 
\noindent In order to test the range for separability as well as inseparability of the output states, we use the Peres-Horodecki criteria \cite{peres}. This is a necessary and sufficient criteria for detecting  entanglement in bipartite systems with dimension $2 \otimes 2$ and $2 \otimes 3$.


\subsection{Broadcasting of entanglement via local cloning}
\label{subsec:broad_ent_local}

\noindent In this subsection, we describe the principle of broadcasting of quantum entanglement by using local cloning transformation. For this we start with a two qubit state $\rho_{12}$ (given in Eq.~\eqref{eq:mix}) shared between two parties $A$ and $B$. The first qubit `1' belongs to the party $A$ while the second qubit `2' belongs to $B$. Each of them will now individually apply the local copying transformation (given by Eq.~\eqref{eq:localcloner}),  on their own qubit to produce the state $\tilde{\rho}_{1234}$ \cite{buzek2, buzek3, sourav}. It is important to note that for all local cloning operations we consider $N=2$, that is the B-H optimal local cloner, since in this operational technique for higher values of $N$ the output nonlocal pairs always remain disentangled \cite{kar}. More formally, it can be defined as follows:\\


\noindent \textbf{Definition 2.1 \cite{buzek2, adhikari, sourav}:} 
\label{def:broad_local}
An entangled state $\rho_{12}$ is said to be broadcast after the application of local cloning operation $U^{1}_{1,2} \otimes U^{2}_{1,2}$, each of the type given by Eq.~\eqref{eq:localcloner}, on the qubits $1$ and $2$ respectively, if for some values of the input state parameters, 
\begin{itemize} 
\item the non-local output states between $A$ and $B$
\begin{eqnarray}
&& \tilde{\rho}_{14} = Tr_{23}\left[U^{1}_{1,2}\otimes U^{2}_{1,2}\left(\rho_{12}\right)\right] \nonumber\\
&& \tilde{\rho}_{23} = Tr_{14}\left[U^{1}_{1,2}\otimes U^{2}_{1,2} \left(\rho_{12}\right) \right]\:\: \text{are inseparable,}
\label{eq:gen_nonlocal_output_localcloner}
\end{eqnarray}
\item whereas the local output states for each of two parties A and B
\begin{eqnarray}
&& \tilde{\rho}_{13} = Tr_{24}\left[U^{1}_{1,2}\otimes U^{2}_{1,2}\left(\rho_{12}\right)\right], \nonumber\\
&& \tilde{\rho}_{24} = Tr_{13}\left[U^{1}_{1,2}\otimes U^{2}_{1,2}\left(\rho_{12}\right)\right] \:\:\text{are separable.}
\label{eq:gen_local_output_localcloner}
\end{eqnarray}
\end{itemize}

\noindent The state  $\rho_{12}$ given in Eq.~\eqref{eq:mix} is a general mixed state and will not be entangled for all values of its parameters. However, when we talk about broadcasting of entanglement it is only relevant to consider those values when the initial state $\rho_{12}$ is entangled. Then the range of input state parameters for which broadcasting will be possible is always going to be a subset of the range of the input state parameters for which $\rho_{12}$ is entangled.\\

\subsection{Broadcasting of entanglement via nonlocal cloning}
\label{subsec:broad_ent_nonlocal}

\noindent In this subsection, we reconsider the problem of broadcasting of entanglement, however this time we use nonlocal cloning transformation to create more pairs. This situation is quite analogous to the previous case where we have used local cloning operations. Here, the basic idea is that the starting state is again $\rho_{12}$ (given in Eq.~\eqref{eq:mix}) and we want to create more entangled copies but with a global unitary operation $U^{12}_{1,N}$ to produce $\tilde{\rho}_{1234}$ \cite{buzek3, sourav}. Importantly, here $N$ can be greater than $2$ unlike the previous case of local cloning, since for $N\leqslant6$ the output pairs have some residual entanglement \cite{kar}. More particularly we can define it as follows: 


\noindent \textbf{Definition 2.2 \cite{buzek2, adhikari}:} An entangled state $\rho_{12}$ is said to be broadcast after the application of nonlocal cloning operation $U^{12}_{1,N}$ (given by Eq.~\eqref{eq:nonlocalcloner}) together on the 
qubits $1$ and $2$, if for some values of the input state parameters,
\begin{itemize} 
\item the desired output states \\
$\tilde{\rho}_{12}$ = $Tr_{34}\left[U^{12}_{1,N}\left(\rho_{12}\right)\right]$,\\
$\tilde{\rho}_{34}$ = $Tr_{12}\left[U^{12}_{1,N}\left(\rho_{12}\right)\right]$ are inseparable,
\item and the remaining output states\\
$\tilde{\rho}_{13}$ = $Tr_{24}\left[U^{12}_{1,N}\left(\rho_{12}\right)\right]$,\\
$\tilde{\rho}_{24}$ = $Tr_{13}\left[U^{12}_{1,N}\left(\rho_{12}\right)\right]$ are separable.
\end{itemize}

\noindent We could have also chosen the diagonal pairs ($\tilde{\rho}_{14}\: \&\: \tilde{\rho}_{23}$) instead of choosing the pairs: $\tilde{\rho}_{12}\: \&\: \tilde{\rho}_{34}$ as our desired pairs in the above definition. However, we refrain ourselves from choosing the pairs $\tilde{\rho}_{13}\: \&\: \tilde{\rho}_{24}$ as the desired pairs \cite{buzek3}.\\

\section{Complementarity of Information Processing Tasks with Broadcasting Fidelity}

\noindent In this section, we present the central idea of our work where we establish the complimentary relations between the fidelity of broadcasting process with the decremental change in the information processing abilities as a consequence of generation of lesser entangled pairs from an initially more entangled resource. As explained in the previous section, this entire process of broadcasting can happen by the use of either local or nonlocal cloning operations. In our complementarity analysis, we consider two of the most common quantum information processing protocols, namely teleportation and superdense coding \cite{ben2, ben3}. However, we conjecture that this complementary nature between broadcasting fidelity and information processing ability of an input entangled state will hold true for all known information processing protocols in the quantum world. After applying either of the cloning process, we separately calculate the change in the maximal teleportation fidelity and the superdense coding capacity of the entangled state to find that these values can not be arbitrary values \cite{horo, chak1}. This change has a trade off with the broadcasting fidelity of the process. In fact the sum of these two quantities is a constant expression. Interestingly this shows us that each of these quantities are complimentary in nature. In other words, better is the fidelity of broadcasting lesser is the change in the information processing capabilities. In short, if we broadcast well we preserve the capacity of the entangled state to be used as an useful resource. Next, we briefly define \\

\noindent \textbf{Teleportation Fidelity: }\\

\noindent Quantum teleportation is sending the quantum information belonging to one party to a distant party with the help of a resource entangled state \cite{ben2}. It is well known that all pure entangled states $2 \otimes 2$ dimensions are useful for teleportation \cite{pati,horo}. However, the situation is not so trivial for mixed entangled states.  There are also entangled states which can not be used as a resource for teleportation \cite{horo}. However after suitable local operation and classical communication (LOCC) one can always convert them to states which become useful for teleportation \cite{vers}. The extent to which a two qubit state can be used as a resource for teleportation is quantified by the maximal fidelity of teleportation ($TF$) \cite{horo}. For a general two qubit mixed state (given by Eq.~\eqref{eq:mix}) as a resource, we have the $TF$ defined for it as, 
\begin{eqnarray}
TF (\rho_{12})=\frac{1}{2}[1+\frac{1}{3}(\sum_i \sqrt{u_i})].
\end{eqnarray}

\noindent where $u_i$ are the eigenvalues of the matrix $U=T^{\dagger}T$. A quantum state is said to be useful for teleportation when $TF$ is more than $\frac{2}{3}$ which is the classically achievable limit of fidelity of teleportation. One such example of an useful resources is the Werner state \cite{werner,wernerlike} in $2 \otimes 2$ dimensions for a certain range of its classical probabilities of mixing \cite{horo}. Other examples, of mixed entangled states as a resource for teleportation also exists \cite{chak}.\\

\noindent \textbf{Super Dense Coding Capacity }\\

\noindent Quantum super dense coding involves in sending of classical information from one sender to the receiver when they are sharing a quantum resource in form of an entangled state. More specifically, superdense coding is a technique used in quantum information theory to transmit classical information by sending quantum systems \cite{ben3}. It is quite well known that if we have a maximally entangled state in $H_d\otimes H_d$ as our resource, then we can send $2 \log d$ bits of classical information. In the asymptotic case, we know one can send $\log d + S(\rho)$ amount of bit when one considers non-maximally entangled state as resource \cite{pank1,bru, ari, shad}. It had been seen that the number of classical bits one
can transmit using a non-maximally entangled state in $H_d\otimes H_d$ as a resource is $(1 + p_0\frac{d}{d-1}) \log d$, 
where $p_0$ is the smallest Schmidt coefficient. However, when the state is maximally entangled in its subspace then one can send up to $2*\log(d − 1)$ bits \cite{chak1}.\\

\subsection{Complementarity for Two qubit generalized mixed states}

\noindent In this subsection, we consider the most generalized two qubit mixed state $\rho_{12}$ as our initial resource given by Eq.~\eqref{eq:mix}. Then we apply generalized Buzek Hillery cloning transformation on this state to obtain four qubit states $\tilde{\rho}_{1234}$ as output. The cloning is carried out both locally on individual qubits and nonlocally on both the qubits. Then in each case we trace out the redundant qubits to obtain the newly generated entangled pairs $\sigma=\tilde{\rho}_{14}/\tilde{\rho}_{23}=\left\{\frac{2}{3}\vec{x}, \frac{2}{3}\vec{y}, \frac{4}{9}T \right\}$ (for local cloning) and $\sigma=\tilde{\rho}_{12}/\tilde{\rho}_{34}= \left\{\frac{3}{5}\vec{x}, \frac{3}{5}\vec{y}, \frac{3}{5}T \right\}$ (for nonlocal cloning). For each of them, we compute the broadcasting fidelity which is given by, 

\begin{eqnarray}
FB(\rho_{12}, \sigma)=tr[\sqrt{\sqrt{\rho_{12}}\sigma\sqrt{\rho_{12}}}].
\label{eq:broadfid}
\end{eqnarray}
We also compute the change in the teleportation fidelity given by,   
\begin{eqnarray}
\Delta TF(\rho_{12}, \sigma)=TF\left(\rho_{12}\right)-TF\left(\sigma\right),
\label{eq:diffTF}
\end{eqnarray}
and the change in the super dense coding capacity given by, 
\begin{eqnarray}
\Delta DC(\rho_{12}, \sigma)=DC\left(\rho_{12}\right)-DC\left(\sigma\right)
\label{eq:diffDC}
\end{eqnarray}

\noindent in the transition of the initial state $\rho_{12}$ to the final state $\sigma$ through the decompression process. Interestingly, we observe that the sum of both these 
quantities with the corresponding broadcasting fidelities is always bounded by a quantity depending on the initial state parameters. We plot these 
sums in subsequent figures for both local and nonlocal cloning transformations. In other words we show that these two quantities namely broadcasting fidelity ($FB$) and the change in these information processing capabilities ($\Delta DC(TF)$) are complimentary to each other. More particularly, an increase in one quantity will bring down the other. \\

\noindent \textit{With Local Cloner:}\\

\begin{figure}[h]
\centering
\subfigure{\includegraphics[scale=0.60]{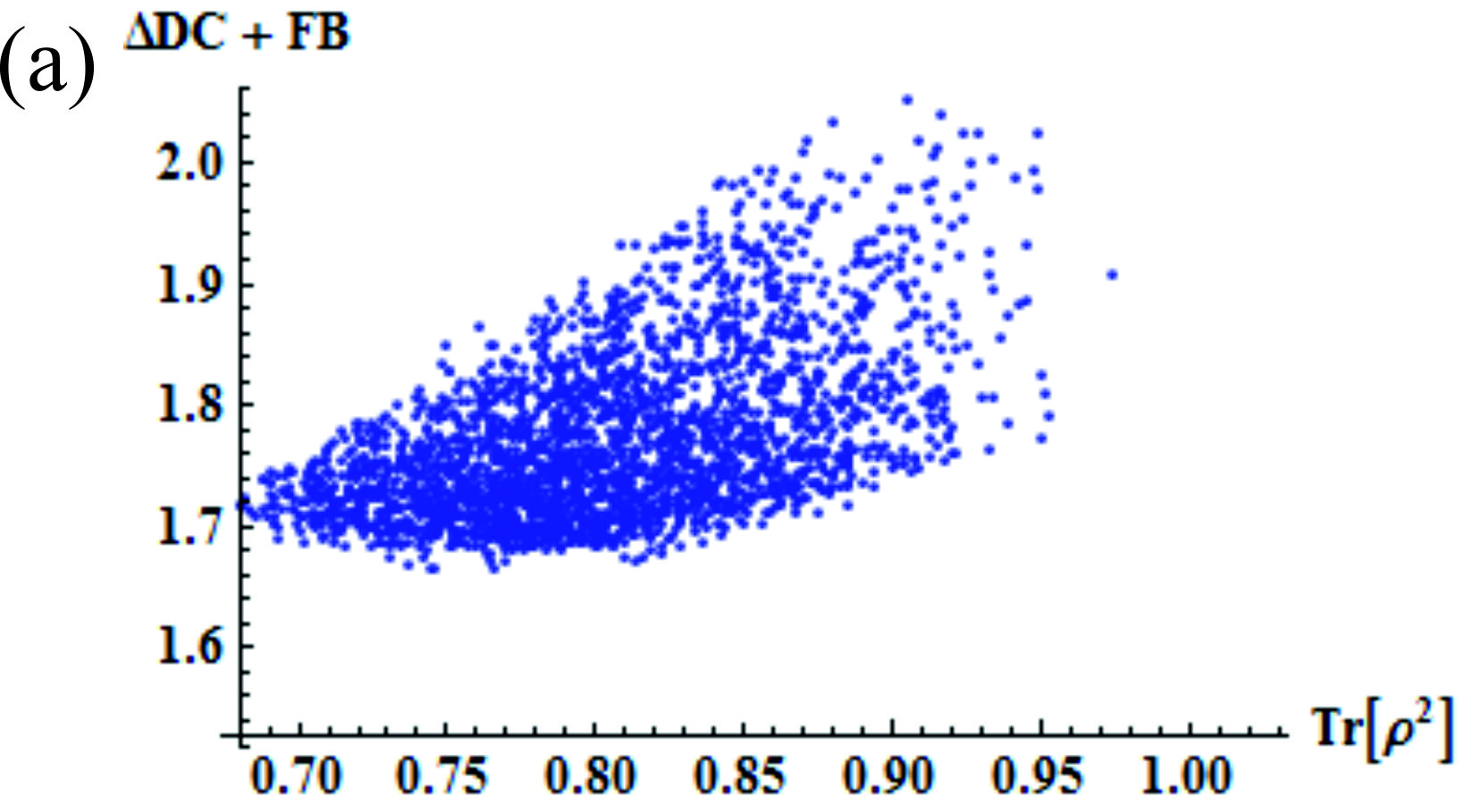}}
\subfigure{\includegraphics[scale=0.60]{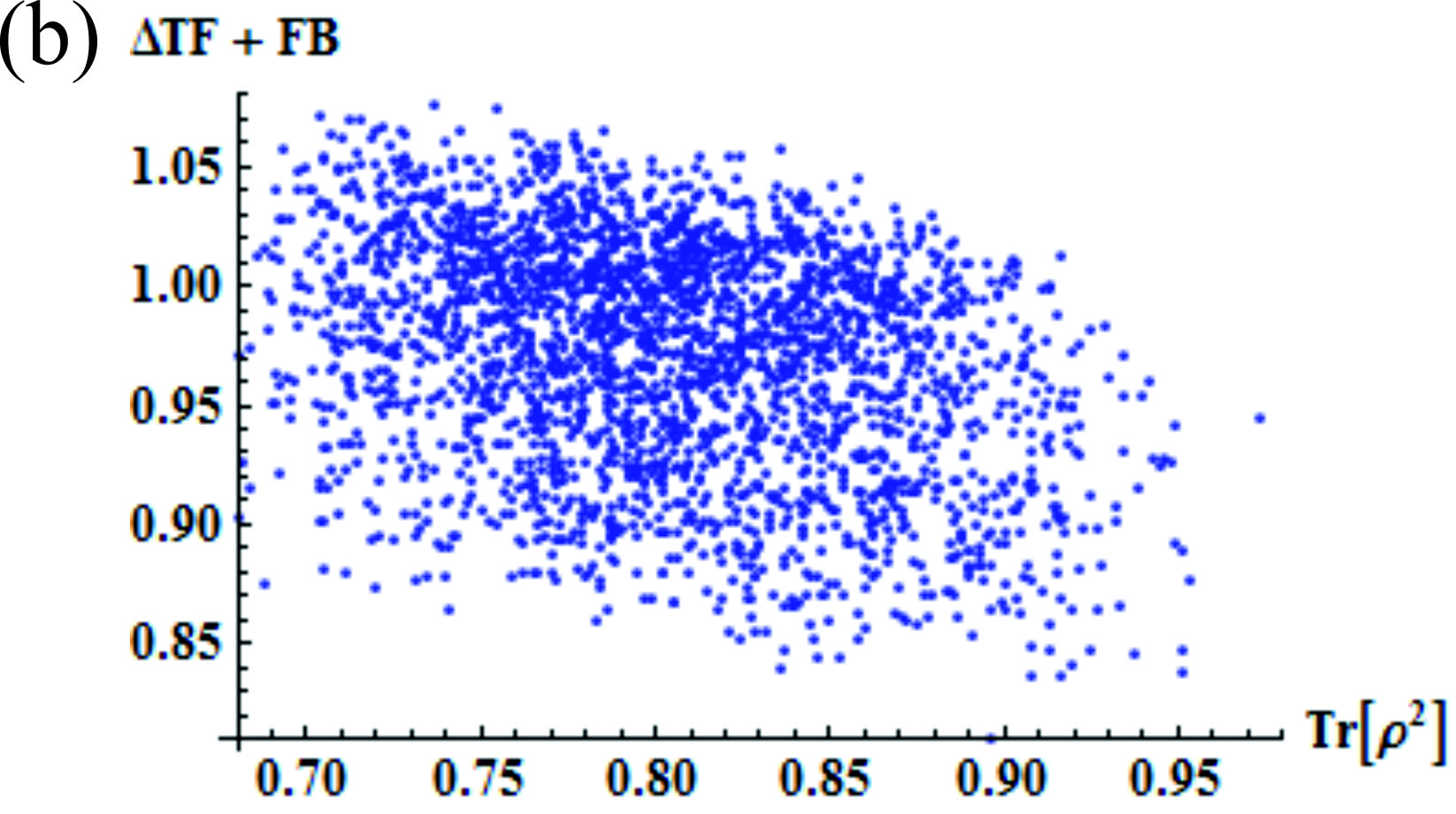}}
\caption{\noindent \scriptsize For local cloning operation: (a) Variation of $\Delta DC+ FB$ bound with $Tr[\rho^2]$ and (b) Variation of $\Delta TF+ FB$ bound with $Tr[\rho^2]$.}
\label{fig:mostgenlocal}
\end{figure}

\noindent In Fig.~\ref{fig:mostgenlocal} we plot the sum a) $\Delta DC+ FB$ and b) $\Delta TF+ FB$ with the trace of the square of the initial state $\rho$. The randomly generated state parameters are set of points on Bloch sphere. Each point represents a two qubit state. This figure corresponds to the situation when we have used local cloning transformation on respective qubits. In both of these cases the respective sums are bounded. In (a) the sum of these quantities can never go beyond $2$ as $FB$ is bounded by $1$ while the teleportation fidelity can never be more than $1$. Similarly in (b) the maximum dense coding capacity for a two qubit state is $2$ (for Bell states) so here $3$ serves as an upper bound to this sum. As for a given state the sum is always fixed, we conclude that an increase in any one of the quantities will bring down the other.\\

\noindent \textit{With Nonlocal Cloner:}\\

\begin{figure}[h]
\centering
\subfigure{\includegraphics[scale=0.60]{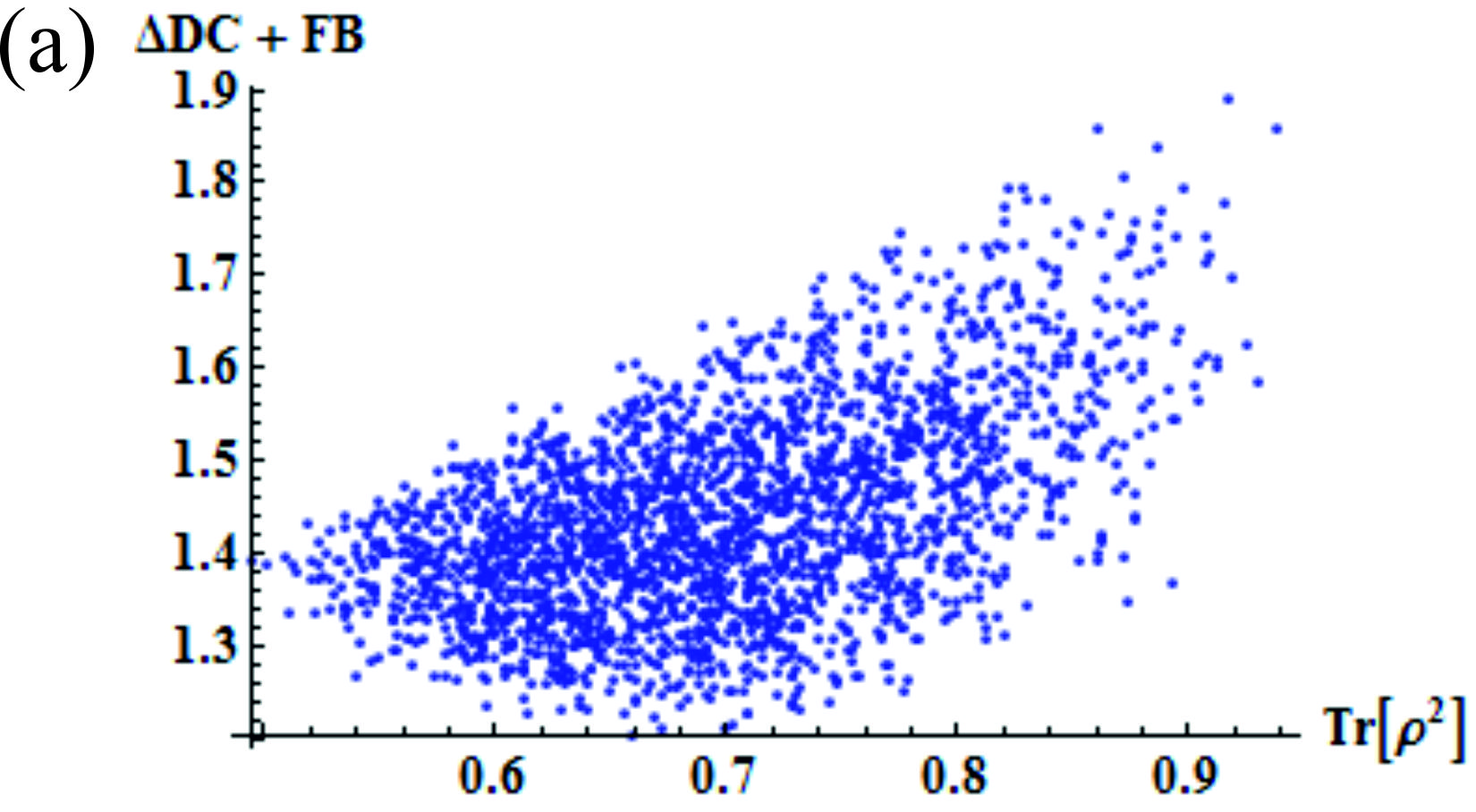}}
\subfigure{\includegraphics[scale=0.60]{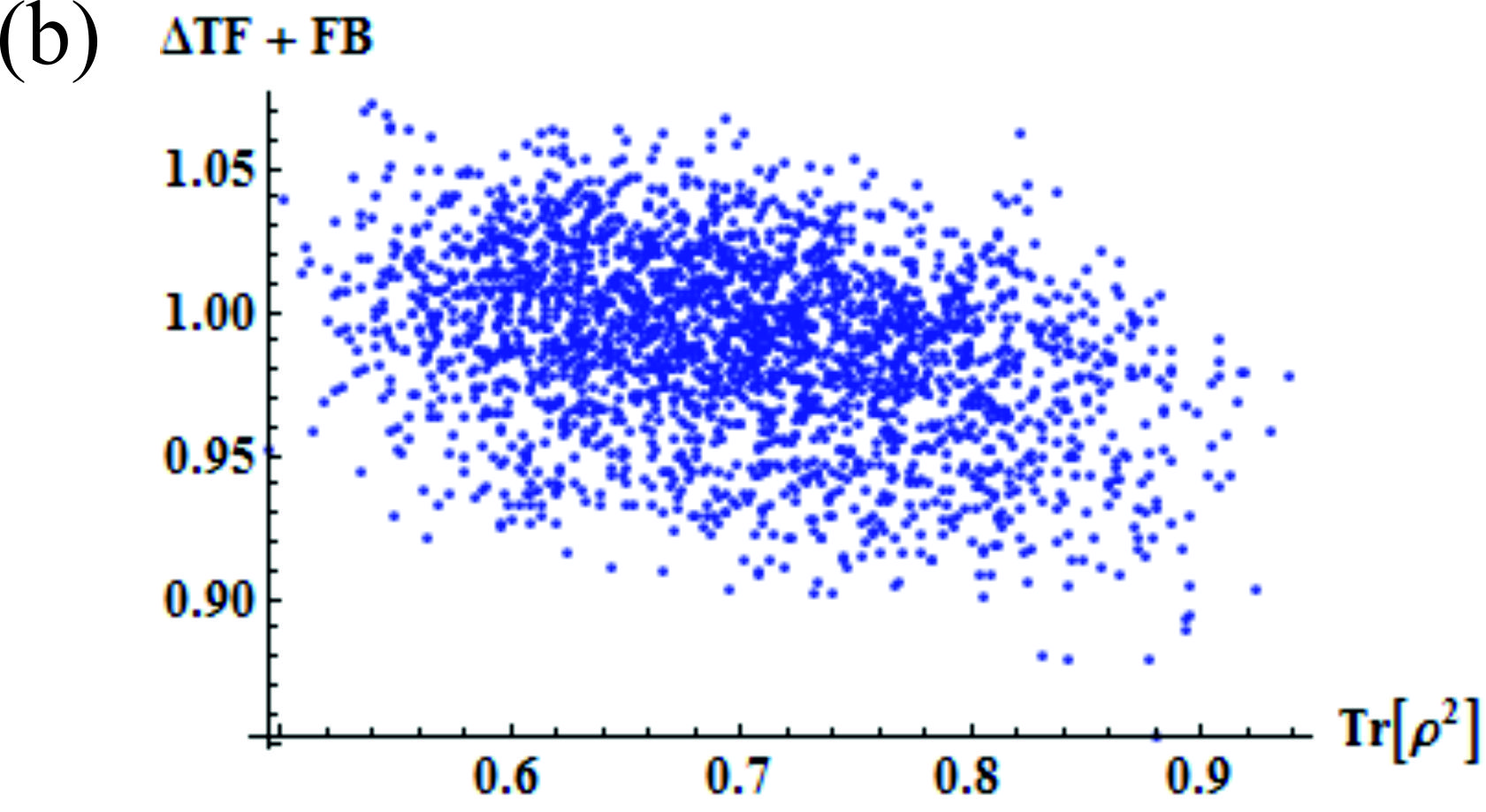}}
\caption{\noindent \scriptsize For nonlocal cloning operation: (a) Variation of $\Delta DC+ FB$ bound with $Tr[\rho^2]$ and (b) Variation of $\Delta TF+ FB$ bound with $Tr[\rho^2]$.}
\label{fig:mostgennonlocal}
\end{figure}

\noindent In Fig.~\ref{fig:mostgennonlocal} we plot the sum a) $\Delta DC+ FB$ and b) $\Delta TF+ FB$ with the trace of the square of the initial state $\rho$. However, in this situation we have used non local cloning transformations for the purpose of broadcasting. Quite similar to the previous figure here also we observe that all these sums are respectively bounded by $2$ and $3$. \\

\noindent Next we exemplify this complementarity with the help of two well known class of mixed states namely, Werner-like and Bell diagonal States. In each of these cases we show that how broadcasting fidelity and the change in the information processing capabilities maintains a complementarity relationship with each other.\\

\noindent \textbf{\textit{Example A: Werner-like States}}\\

 \noindent \textbf{Broadcasting via Local Cloning:}\\

\noindent In this example, we consider the class of Werner-like states \cite{wernerlike} as our resource states. These states can more formally be expressed as, $\rho^w_{12} = \left\{\vec{x}^w, \vec{x}^w, T^w \right\}$, where $\vec{x}^w = \left\{ 0,\:  0,\:  p \left(\alpha^2-\beta^2\right)\right\}$ is the Bloch vector and the correlation matrix $T^w=\left(\begin{smallmatrix} 2p\alpha\beta&0&0\\ 0&-2p\alpha\beta&0\\0&0&p\end{smallmatrix} \right)$ with the condition $\alpha^2+\beta^2=1$.\\



\noindent After applying the optimal local cloning process given by Eq.~\eqref{eq:localcloner}, the nonlocal output states appear to be, $\tilde{\rho}_{14}=\tilde{\rho}_{23}= \left\{\frac{2}{3}\vec{x}^w,\: \frac{2}{3}\vec{x}^w,\: \frac{4}{9}T^w \right\}$. Using Peres-Horodecki Theorem, we discover that the broadcasting range is given by \cite{sourav},
\begin{equation}
\frac{3}{4}<p\leq1\, \text{ \& }\,N_-<\alpha^2 < N_+,
\label{eq:werner_broadrange_local}
\end{equation}
where $N_{\pm}=\frac{1}{16}\{8\pm(48-\frac{81}{p^2}+\frac{72}{p})^{\frac{1}{2}}\}$. \\


\noindent Next we provide two different tables for detailed analysis of the above broadcasting range. In TABLE~\ref{tbl:werner_like_local_1}, we give the broadcasting range of the werner-like states in terms $p$ for the different values of the input state parameter $\alpha^2$. Here, we also calculate the decremental effect caused to the maximal teleportation fidelity $\Delta TF$ and superdense coding capacity $\Delta DC$ as a result of the broadcasting process. The sum of each of these quantities with broadcasting fidelity $FB$ for a given value of $\alpha^2$ are provided in this table. There is a clear indication that sum of these quantities is constant for a given value of the input state parameters. \\
\begin{center}
\begin{table}[!ht]
\begin{tabular}{|c|c|c|c|}
\hline 
$\alpha^{2}$ & Broadcasting & $(\Delta TF + FB)$  & $(\Delta DC + FB)$\tabularnewline
 & Range& &\tabularnewline
\hline 
\hline 
$0.2$ & $0.87<p\leqslant1$ & $1.061$ & $2.017$ \tabularnewline
\hline 
$0.4$ & $0.76<p\leqslant1$ & $1.096$ & $2.221$ \tabularnewline
\hline
$0.5$ & $0.75<p\leqslant1$ & $1.099$ & $2.244$\tabularnewline
\hline 
$0.6$ & $0.76<p\leqslant1$ & $1.096$ & $2.221$ \tabularnewline
\hline 
$0.8$ & $0.87<p\leqslant1$ & $1.061$ & $2.017$\tabularnewline
\hline 
\end{tabular}
\caption{\noindent \scriptsize Broadcasting range and complementarity bounds obtained using local cloners for different values of $\alpha^2$.}
\label{tbl:werner_like_local_1}
\end{table}
\end{center}

\noindent In TABLE~\ref{tbl:werner_like_local_2}, we give the broadcasting range in terms of $\alpha^2$ for different values of the classical mixing parameter $p$. Quite similar to the previous table, here also we give the values for the sums $(\Delta TF + FB)$  and $(\Delta DC + FB)$ against the various values of the classical randomness parameter $p$.\\

\begin{center}
\begin{table}[!ht]
\begin{tabular}{|c|c|c|c|}
\hline 
$p$  & Broadcasting & $(\Delta TF + FB)$  & $(\Delta DC + FB)$ \tabularnewline
& Range & &\tabularnewline
\hline 
\hline 
$0.76$ & $0.40<\alpha^2<0.60$ & $1.097$ & $1.708$ \tabularnewline
\hline 
$0.8$ & $0.29<\alpha^2<0.71$ & $1.091$ & $1.786$\tabularnewline
\hline
$0.85$ & $0.22<\alpha^2<0.78$ & $1.077$ & $1.892$\tabularnewline
\hline 
$0.9$ & $0.17<\alpha^2<0.83$ & $1.055$ & $2.007$\tabularnewline
\hline 
$0.95$ & $0.14<\alpha^2<0.87$ & $1.014$ & $2.134$  \tabularnewline
\hline 
$1$ & $0.11< \alpha^2 <0.89$ & $0.864$ & $2.224$\tabularnewline
\hline 
\end{tabular}
\caption{\noindent \scriptsize Broadcasting range and complementarity bounds obtained with local cloners for different values of $p$.}
\label{tbl:werner_like_local_2}
\end{table}
\end{center}

\begin{figure}[h]
\centering
\subfigure{\includegraphics[scale=0.70]{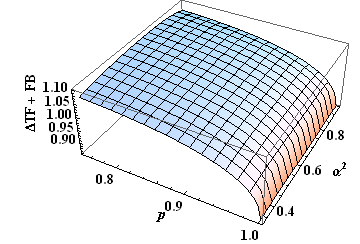}}
\subfigure{\includegraphics[scale=0.70]{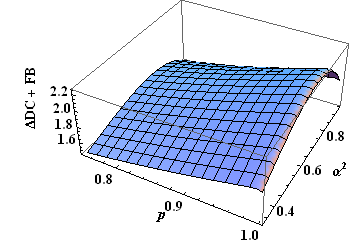}}
\caption{\noindent \scriptsize Plotting $(\Delta TF + FB)$ and $(\Delta DC + FB)$ against the parameters $\alpha^2$ and $p$ of the Werner-like state for local cloning operation}.
\label{fig:wernerlike_local}
\end{figure}

\noindent Next we plot each of these sums in FIG.~\ref{fig:wernerlike_local} against two state parameters $\alpha^2$ and $p$. We find that each of these sums $(\Delta TF + FB)$ and $(\Delta DC + FB)$ are bounded establishing once again the mutually exclusive nature of both these quantities  $\Delta TF(DC)$ and $FB$.\\

\noindent \textbf{Broadcasting by Non Local Cloning:}\\

\noindent After applying the nonlocal cloning operation Eq.~\eqref{eq:nonlocalcloner}, the desired output states in this case are given by, $\tilde{\rho}_{12}=\tilde{\rho}_{34}= \left\{\frac{3}{5}\vec{x}^w,\: \frac{3}{5}\vec{x}^w,\: \frac{3}{5}T^w \right\}$, where, $\vec{x}^w$ is the Bloch vector and $T^w$ is the correlation matrix of werner-like state. The remaining output states are given by,
$\tilde{\rho}_{13}=\tilde{\rho}_{24}=\left\{\frac{3}{5}\vec{x}^w,\: \frac{3}{5}\vec{x}^w,\: \frac{1}{5}\mathbb{I}_3 \right\}$.\\

\noindent The broadcasting (inseparability) range of the desired output states is given by \cite{sourav},
$\frac{5}{9}<p\leqslant1\, \text{ and }\,\frac{1}{2}-H<\alpha^2 < \frac{1}{2}+H,$
where $H=\sqrt{\frac{27p^2+30p-25}{144p}}$.\\

\noindent Quite similar to the local cloning situation here also we provide two different tables for detailed analysis of 
the broadcasting range. In TABLE~\ref{tbl:werner_like_nonlocal_1}, we give the broadcasting range in terms of the 
classical mixing parameter $p$ for given values of input state parameter $\alpha^2$. In this table, we have also given the values of $(\Delta TF + FB)$  and $(\Delta DC + FB)$ for the same set of values of the input state parameters.\\

\begin{center}
\begin{table}[!ht]
\begin{tabular}{|c|c|c|c|}
\hline 
$\alpha^{2}$ & Broadcasting  & $(\Delta TF + FB) $  & $(\Delta DC + FB) $\tabularnewline
&    Range & &\tabularnewline
\hline 
\hline 
$0.2$ & $0.64<p\leqslant1$ & $1.071$ & $1.894$\tabularnewline
\hline 
$0.4$ & $0.56<p\leqslant1$ & $1.088$ & $2.057$\tabularnewline
\hline
$0.5$ & $0.55<p\leqslant1$ & $1.090$ & $2.075$\tabularnewline
\hline 
$0.6$ & $0.56<p\leqslant1$ & $1.088$ & $2.057$\tabularnewline
\hline 
$0.8$ & $0.64<p\leqslant1$  & $1.071$ & $1.894$\tabularnewline
\hline 
\end{tabular}
\caption{\noindent \scriptsize Broadcasting range and complementarity bounds obtained with a nonlocal cloner for different values of the input state parameter ($\alpha^2$).}
\label{tbl:werner_like_nonlocal_1}
\end{table}
\end{center}

\noindent In TABLE~\ref{tbl:werner_like_nonlocal_2}, we give the range of broadcasting in terms of the input state 
parameter $\alpha^2$ for given values of classical mixing parameter $p$ and for same values of $p$ we have also provided the values of the sums $(\Delta TF + FB) $  and $(\Delta DC + FB)$.

\begin{center}
\begin{table}[!ht]
\begin{tabular}{|c|c|c|c|}
\hline 
 $p$  & Broadcasting & $(\Delta TF + FB) $  & $(\Delta DC + FB) $\tabularnewline
 & Range & &\tabularnewline
\hline 
\hline 
$0.56$ & $0.42<\alpha^2<0.58$ & $1.083$ & $1.322$\tabularnewline
\hline 
$0.65$ & $0.19<\alpha^2<0.81$ & $1.089$ & $1.432$\tabularnewline
\hline
$0.75$ & $0.10<\alpha^2<0.90$ & $1.089$ & $1.577$\tabularnewline
\hline
$0.85$ & $0.06<\alpha^2<0.94$ & $1.077$ & $1.753$ \tabularnewline
\hline 
$0.95$ & $0.04< \alpha^2< 0.96$ & $1.032$ & $1.970$ \tabularnewline
\hline 
$1$ & $0.03< \alpha^2 < 0.97$ & $0.9$ & $2.057$\tabularnewline
\hline 
\end{tabular}
\caption{\noindent \scriptsize Broadcasting range and complementarity bounds obtained with a nonlocal cloner for different values of the classical mixing parameter ($p$).}
\label{tbl:werner_like_nonlocal_2}
\end{table}
\end{center}

\begin{figure}[h]
\centering
\subfigure{\includegraphics[scale=0.70]{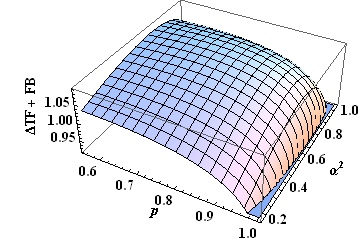}}
\subfigure{\includegraphics[scale=0.70]{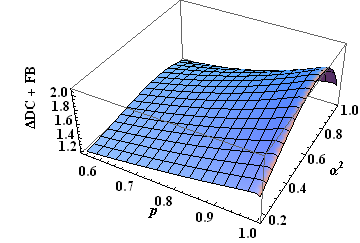}}
\caption{\noindent \scriptsize Plotting $(\Delta TF + FB)$ and $(\Delta DC + FB)$ against the parameters $\alpha^2$ and $p$ of the Werner-like state for nonlocal cloning operation}.
\label{fig:wernerlike_nonlocal}
\end{figure}

\noindent Similar to local cloning case, here also we plot each of these sums in FIG.~\ref{fig:wernerlike_nonlocal} against two state parameters $\alpha^2$ and $p$. We again find that each of these sums $(\Delta TF + FB)$ and $(\Delta DC + FB)$ are bounded re-establishing the mutually exclusive nature of both these quantities  $\Delta TF(DC)$ and $FB$.

\noindent \textbf{\textit{Example B: Bell Diagonal States}}\\

\noindent Next we consider the example of Bell diagonal states \cite{belld} as our resource. The input bell-diagonal state to the cloners can be more formally expressed as, $\rho^b_{12} = \left\{\vec{0}, \vec{0}, T^b\right\}$, where $\vec{0}$ is the Bloch vector which is a null vector and the correlation matrix 
$T^b=\left(\begin{smallmatrix}
c_1&0&0\\ 0&c_2&0\\0&0&c_3 
\end{smallmatrix} \right)$ with $-1\leqslant \text{c}_{i} \leqslant1$. The input bell-diagonal state can be rewritten as 
$\rho^b_{12}=\sum_{m,n} \lambda_{mn} \left|\gamma_{mn}\right\rangle\left\langle\gamma_{mn}\right|$ where the four Bell states  $\left|\gamma_{mn}\right\rangle\equiv\left(\left|0,n\right\rangle+(-1)^m \left|1,1\oplus n \right\rangle\right)/\sqrt{2}$ represents the eigenstates of $\rho^b_{12}$ with eigenvalues, $\lambda_{mn}=\frac{1}{4}\left[ 1+ (-1)^m c_1-(-1)^{(m+n)} c_2 + (-1)^n c_3\right]$. Also, for $\rho^b_{12}$ to be a valid density operator, its eigenvalues have to be positive, i.e. $\lambda_{mn}\geqslant 0$.\\

\noindent \textbf{Broadcasting by Local Cloning:}\\

\noindent Now by applying local cloning operation (given by Eq.~\eqref{eq:localcloner}) on the input state $\rho^b_{12}$ and then tracing out the qubits we get the local output states as:
$\tilde{\rho}_{13}=\tilde{\rho}_{24}=\left\{\vec{0}, \vec{0}, \frac{1}{3}\mathbb{I}_{3}\right\}$. It turns out that for these local output states $\tilde{\rho}_{13}$ and $\tilde{\rho}_{24}$,  both $W_3$ and $W_4$  are non-negative and are independent of the input state parameters ($c_i$'s). Hence, these local output states always remain separable \cite{sourav}.

On the other hand, the nonlocal outputs are given by, 
$\tilde{\rho}_{14}=\tilde{\rho}_{23}=\left\{\vec{0}, \vec{0}, \frac{4}{9}T^b\right\},$ where $T^b$ is the correlation matrix of the state.\\

\noindent The broadcasting (inseparability) range for these nonlocal output states of the input bell-diagonal 
state $\rho^b_{12}$ in terms of $c_i$'s, is given by \cite{sourav},
\begin{eqnarray}
&& \left(-1\leq c_1 <-\frac{1}{4}\: \text{and}\: \left(c_1+c_2+c_3<-\frac{9}{4}  \: \text{or}\:  c_1-c_3\right. \right. \nonumber\\
&&\left. \left. +\frac{9}{4}  <c_2\leq 1\right)\right)\: \text{or}\: \left(\frac{1}{4}<c_1\leq 1\: \text{and}\:  \left(\frac{9}{4}-c_1\right. \right. \nonumber\\
&& \left.\left. +c_3<c_2 \leq 1 \:\text{or}\: -1\leq c_2<c_1+c_3-\frac{9}{4}\right)\right),
\label{eq:bell_broadrange_local}
\end{eqnarray}
along with the condition that $\lambda_{mn} \geqslant 0$. Since the local output states in this case are always separable, 
it is evident that the broadcasting range of the bell-diagonal state is same as the inseparability range.\\

\noindent In the TABLE~\ref{tbl:bell_diagonal_local}, we give the broadcasting range of Bell-diagonal states $\rho_{12}$ for different values of the first two input state parameters $c_1$, $c_2$. The range is given over the third input state parameter $c_3$, between the valid zone from $-1$ to $-5/8$ or $\frac{5}{8}$ to $1$. Beside the broadcasting range we also provide the maximum values of the sums  $(\Delta TF + FB) max$ and $(\Delta DC + FB) max$  for the corresponding input 
parameter $c_3$. In this table, we restrict our results only to the negative range of inputs for $c_1$ and $c_2$. The broadcasting range in terms of $c_3$ will remains unchanged when corresponding positive values of $c_1$ and $c_2$ are substituted in Eq.~\eqref{eq:bell_broadrange_local}.\\

\begin{table}[h]
\small
\begin{tabular}{| c | c | c | c | c|}
\hline
$c1$  &  $c2$ & c3 & $(\Delta TF + FB) max$  & $(\Delta DC + FB) max$ \\
\hline
$\frac{-7}{8}$  &  $\frac{-7}{8}$ & $-1 \leq c3 \leq \frac{-3}{4}$ & $1.070 $  & $2.054$ \\
\hline
$\frac{-3}{4}$  &  $\frac{-3}{4}$ & $-1 \leq c3 < \frac{-3}{4}$ & $1.099$  & $1.911$ \\
\hline
$\frac{-7}{8}$  &  $\frac{-3}{4}$ & $\frac{-7}{8} \leq c3 < \frac{-5}{8}$ & $1.079$  & $1.866$ \\
\hline
$\frac{-3}{4}$  &  $\frac{-7}{8}$ & $\frac{-7}{8} \leq c3 < \frac{-5}{8}$ & $1.079$  & $1.866$ \\
\hline
\end{tabular}
\caption{\noindent \scriptsize Broadcasting range and complementarity bounds on $(\Delta TF + FB) max$ and $(\Delta DC + FB) max$ obtained with local cloners for different valid values of input state parameters $c_1$ and $c_2$.}
\label{tbl:bell_diagonal_local}
\end{table}

\noindent \textbf{Broadcasting by Non Local Cloning:}\\

\noindent Once the nonlocal cloner is applied to it we have the desired output states as, 
$\tilde{\rho}_{12}=\tilde{\rho}_{34}= \left\{\vec{0}, \vec{0}, \frac{3}{5}T^b\right\}$, where $T^b$ is the the correlation matrix of the state.\\

\noindent The inseparability range of the desired output states is given by \cite{sourav},
\begin{eqnarray}
&& (6c_1-3\gamma +5) (3\gamma-6c_3-5) (3\gamma-6 c_2-5)(3\gamma+ 
 5)<0\:\nonumber\\
 && \text{or}\: (3 c_3+5) \left((5-3 c_3)^2-9 (c_1-c_2)^2\right)<0 \:\:,
\label{eq:bell_broadrange_nonlocal}
\end{eqnarray}
where $\gamma=\tr(T^b)$ along with the condition that $\lambda_{mn} \geqslant 0$ from the positivity of input density operator $\rho_{12}$. The remaining output states are given by, $\tilde{\rho}_{13}\:=\:\tilde{\rho}_{24}= \left\{\vec{0}, \vec{0}, \frac{1}{5}\mathbb{I}_{3}\right\}$.

\noindent These output states are independent of the input state parameter ($c_i$'s) \cite{sourav}. Hence, the broadcasting range of the bell-diagonal state is same as the inseparability range as given in Eq.~\eqref{eq:bell_broadrange_nonlocal} \cite{sourav}.\\

\noindent Hence in TABLE~\ref{tbl:bell_diagonal_nonlocal}, we give the broadcasting range of bell-diagonal states 
$\rho_{12}$ for different values of the first two input state parameters $c_1$, $c_2$. The range is given in terms 
of the  third input state $c_3$, between the valid zone from $-1$ to $-\frac{1}{3}$ or $\frac{1}{3}$ to $1$. In addition, 
to that we have also given the maximum values of the sums  $(\Delta TF + FB) max$ and $(\Delta DC + FB) max$  in that broadcastable zone to show the complimentary nature of these quantities. In this table, we restrict our results only to the negative range of inputs for $c_1$ and $c_2$ . The broadcasting range in terms of $c_3$ remains unchanged when corresponding positive values of $c_1$ and $c_2$ are substituted in Eq.~\eqref{eq:bell_broadrange_nonlocal}.

\begin{table}[h]
\small
\begin{tabular}{| c | c | c | c | c|}
\hline
$c1$  &  $c2$ & $c3$ & $(\Delta TF + FB) max$  & $(\Delta DC + FB) max$ \\
\hline
$\frac{-7}{9}$  &  $\frac{-7}{9}$ & $-1 \leq c3 \leq \frac{-5}{9}$ & $1.088 $  & $1.811$ \\
\hline
$\frac{-5}{9}$  &  $\frac{-5}{9}$ & $-1 \leq c3 < \frac{-5}{9}$ & $1.083$  & $1.665$ \\
\hline
$\frac{-7}{9}$  &  $\frac{-5}{9}$ & $\frac{-7}{9} \leq c3 < \frac{-1}{3}$ & $1.078$  & $1.556$ \\
\hline
$\frac{-5}{9}$  &  $\frac{-7}{9}$ & $\frac{-7}{9} \leq c3 < \frac{-1}{3}$ & $1.078$  & $1.556$ \\
\hline
\end{tabular}
\caption{\noindent  \scriptsize Broadcasting range and complementarity bounds on $(\Delta TF + FB) max$ and $(\Delta DC + FB) max$ obtained with nonlocal cloners for different valid values of input state parameters $c_1$ and $c_2$.}
\label{tbl:bell_diagonal_nonlocal}
\end{table}

\subsection{Complementarity in broadcasting multiple entangled copies}

\noindent In this section, we extend our result to a situation where we have used $1-N$ state independent nonlocal cloning machine (given by Eq.~\eqref{eq:nonlocalcloner}) on our initial resource state $\rho_{12}$ (refer Eq.~\eqref{eq:mix}) to broadcast more than two copies of entanglement ($N>2$ copies). As stated before, with the state independent optimal local cloning machine we can not broadcast more than two copies of the entangled state. However, with the state independent nonlocal cloning we can broadcast upto six copies of entangled state. In this section, we investigate these bounds within the broadcasting range for various cases like $1-2$, $1-3$, $1-4$ and $1-5$ cloning transformations. Instead of showing for general two qubit mixed states, we consider examples like (a) Werner-like state and (b) Bell diagonal state to exhibit the complimentary phenomenon.\\

\noindent \textbf{\textit{Example A: Werner Like states}}:
As a first example, we once again consider the Werner like state $\rho^w_{12} = \left\{\vec{x}^w, \vec{x}^w, T^w \right\}$,
(where $\vec{x}^w = \left\{ 0,\:  0,\:  p \left(\alpha^2-\beta^2\right)\right\}$ is the Bloch vector and $T^b$ is the correlation matrix ) as our initial resource and apply $1-N$ nonlocal cloning machine (given by Eq.~\eqref{eq:nonlocalcloner}) together on both the qubits. In TABLES \ref{wernerlike1} and \ref{wernerlike2} (see Appendices), along with the broadcasting range we also give the bounds on the sum of the broadcasting fidelity ($FB$) and change in maximal teleportation fidelity ($\Delta TF$) for different range of the input state parameters $\alpha^2$ and $p$ respectively. Similarly, in TABLES \ref{wernerlike3} and \ref{wernerlike4} (see Appendices), we repeat the same analysis for the superdense coding capacity ($\Delta DC$) instead of maximal teleportation fidelity. In each of these tables, we give the range as well as the bounds on the sum for $N$ no. of copies, where $N=\{2,3,4,5\}$. \\

\noindent \textbf{\textit{Example B: Bell Diagonal states}}: 
Next we consider the class of Bell diagonal state $\rho^b_{12} = \left\{\vec{0}, \vec{0}, T^b\right\}$ (where $\vec{0}$ is the Bloch vector which is a null vector and $T^b$ is the correlation matrix) as our resource. Here also in TABLES \ref{belldiagonal1} and \ref{belldiagonal2} (see Appendices) we provide the bounds of the sums like $\Delta TF + FB$ and $\Delta DC + FB$ along with the broadcasting range different selected values of the input state parameters respectively. In each of these tables we provide bounds for different copies (where no. of copies, $N=\{2,3,4,5\}$).


\newpage

\begin{widetext}
\section{Appendices}
For all tables below, number of cloned copies of the input state $N =\{2,3,4,5\}$. 
\begin{table}[h]
\begin{tabular}{|p{0.8cm} | p{1.3cm} | p{2.5cm} | p{1.3cm} | p{2.5cm}| p{1.3cm} | p{2.5cm}| p{1.3cm} | p{2.5cm} |}
\hline
No. of copies & \multicolumn{2}{| c |}{2 copies} & \multicolumn{2}{| c |}{3 copies} & \multicolumn{2}{| c |}{4 copies} & \multicolumn{2}{| c |}{5 copies} \\
\hline
\hline
 $\alpha ^ 2$   & $p$ Range & $(\Delta TF + FB)$max & $p$ Range & $(\Delta TF + FB)$max  & $p$ Range & $(\Delta TF + FB)$max  & $p$ Range & $(\Delta TF + FB)$max  \\
\hline
$.2$ & $ .64 - 1$ & $1.071$ & $.82 - 1$ & $1.057 $ & $.96 - 1 $ & $.955$ & \multicolumn{2}{|c|}{$NA$} \\
\hline
$.4$ & $ .56 - 1$ & $1.088$ & $.72 - 1$ & $1.098$ & $.84 - 1 $ & $1.075$ & $.93 - 1$ & $1.018$ \\
\hline
$.5$ & $ .55 - 1$ & $1.090$ & $.71 - 1$ & $1.101$ & $.83 - 1 $ & $1.082$ & $.84 - 1 $ & $1.030$ \\
\hline
$.6$ & $ .56 - 1$ & $1.088$ & $.72 - 1$ & $1.098$ & $.84 - 1 $ & $1.075$ & $.84 - 1 $ & $1.018$\\
\hline
$.8$ & $ .64 - 1$ & $1.071$ & $.82 - 1$ & $1.057$ & $.96 - 1 $ & $.955$ & \multicolumn{2}{|c|}{$NA$} \\
\hline
\end{tabular}
\caption{[Werner Like States] Complementarity bounds for the change in maximal teleportation fidelity and broadcasting fidelity obtained with a $1-N$ nonlocal cloner for different values of the input state parameter($\alpha^2$)}
\label{wernerlike1}
\end{table}

\begin{table}[h]
\begin{tabular}{|p{0.8cm} | p{1.3cm} | p{2.5cm} | p{1.3cm} | p{2.5cm}| p{1.3cm} | p{2.5cm}| p{1.3cm} | p{2.5cm} |}
\hline
No. of copies & \multicolumn{2}{| c |}{2 copies} & \multicolumn{2}{| c |}{3 copies} & \multicolumn{2}{| c |}{4 copies} & \multicolumn{2}{| c |}{5 copies} \\
\hline
\hline
 $p$   & $\alpha ^ 2$ Range & $(\Delta TF + FB)$max & $\alpha ^ 2$ Range & $(\Delta TF + FB)$max  & $\alpha ^ 2$ Range & $(\Delta TF + FB)$max  & $\alpha ^ 2$ Range & $(\Delta TF + FB)$max  \\
\hline
$.56$ & $ .42 - .58$ & $1.083$ & \multicolumn{2}{|c|}{$NA$} & \multicolumn{2}{|c|}{$NA$} & \multicolumn{2}{|c|}{$NA$}   \\
\hline
$.65$ & $ .19 - .81$ & $1.089$ & \multicolumn{2}{|c|}{$NA$} &\multicolumn{2}{|c|}{$NA$}  & \multicolumn{2}{|c|}{$NA$}   \\
\hline
$.75$ & $ .10 - .90 $ & $1.089$ & $.42 - .58$ & $1.098$ &\multicolumn{2}{|c|}{$NA$}  & \multicolumn{2}{|c|}{$NA$}  \\
\hline
$.85$ & $ .06 - .94$ & $1.077$ & $.17 - .83$ & $1.078$ & $.61 - .79$ & $1.071$ & \multicolumn{2}{|c|}{$NA$}  \\
\hline
$.95$ & $ .04 - .96$ & $1.032$ & $.11 - .89$ & $1.017$ & $.45 - .89$ & $1.007$ & $ .60 - .80$ & $.996$ \\
\hline
$1$ & $ .03 - .97$ & $.9$ & $.08 - .92$ & $.867$ & $.41 - .92$ & $.85$ & $ .52 - .85$ & $.840$\\
\hline
\end{tabular}
\caption{[Werner Like States] Complementarity bounds for the change in maximal teleportation fidelity and broadcasting fidelity obtained with a $1-N$ nonlocal cloner for different values of the input state parameter($p$)}
\label{wernerlike2}
\end{table}

\begin{table}[h]
\begin{tabular}{|p{0.8cm} | p{1.3cm} | p{2.5cm} | p{1.3cm} | p{2.5cm}| p{1.3cm} | p{2.5cm}| p{1.3cm} | p{2.5cm} |}
\hline
No. of copies & \multicolumn{2}{| c |}{2 copies} & \multicolumn{2}{| c |}{3 copies} & \multicolumn{2}{| c |}{4 copies} & \multicolumn{2}{| c |}{5 copies} \\
\hline
\hline
 $\alpha ^ 2$   & $p$ Range & $(\Delta DC + FB)$max & $p$ Range & $(\Delta DC + FB)$max  & $p$ Range & $(\Delta DC + FB)$max  & $p$ Range & $(\Delta DC + FB)$max  \\
\hline
$.2$ & $ .64 - 1$ & $1.894$ & $.82 - 1$ & $2.007$ & $.96 - 1$ & $2.043$ & \multicolumn{2}{| c |}{$NA$}  \\
\hline
$.4$ & $ .56 - 1$ & $2.057$ & $.72 - 1$ & $2.203$ & $.84 - 1$ & $2.253$ & $.93 - 1$ & $2.276$\\
\hline
$.5$ & $ .55 - 1$ & $2.075$ & $.71 - 1$ & $2.226$ & $.83 - 1$ & $2.277$ & $.92 - 1$ & $2.301$\\
\hline
$.6$ & $ .56 - 1$ & $2.057$ & $.72 - 1$ & $2.203$ & $.84 - 1$ & $2.253$ & $.93 - 1$ & $2.276$ \\
\hline
$.8$ & $ .64 - 1$ & $1.894$ & $.82 - 1$ & $2.007$ & $.96 - 1$ & $2.043$ & \multicolumn{2}{| c |}{$NA$} \\
\hline
\end{tabular}
\caption{[Werner Like States] Complementarity bounds for the change in superdense coding capacity and broadcasting fidelity obtained 
with a $1-N$ nonlocal cloner for different values of the input state parameter($\alpha^2$)}
\label{wernerlike3}
\end{table}

\begin{table}[!h]
\centering
\begin{tabular}{|p{0.8cm} | p{1.3cm} | p{2.5cm} | p{1.3cm} | p{2.5cm}| p{1.3cm} | p{2.5cm}| p{1.3cm} | p{2.5cm} |}
\hline
No. of copies & \multicolumn{2}{| c |}{2 copies} & \multicolumn{2}{| c |}{3 copies} & \multicolumn{2}{| c |}{4 copies} & \multicolumn{2}{| c |}{5 copies} \\
\hline
\hline
 $p$   & $\alpha ^ 2$ Range & $(\Delta DC + FB)$max & $\alpha ^ 2$ Range & $(\Delta DC + FB)$max  & $\alpha ^ 2$ Range & $(\Delta DC + FB)$max  & $\alpha ^ 2$ Range & $(\Delta DC + FB)$max  \\
\hline
$.56$ & $ .42 - .58$ & $1.322$ & \multicolumn{2}{|c|}{$NA$} & \multicolumn{2}{|c|}{$NA$} & \multicolumn{2}{|c|}{$NA$}\\
\hline
$.65$ & $ .19 - .81$ & $1.432$ & \multicolumn{2}{|c|}{$NA$}& \multicolumn{2}{|c|}{$NA$} & \multicolumn{2}{|c|}{$NA$}\\
\hline
$.75$ & $ .10 - .90$ & $1.577$ & $.42 -  .58$ & $1.677$ & \multicolumn{2}{|c|}{$NA$} & \multicolumn{2}{|c|}{$NA$}\\
\hline
$.85$ & $ .06 - .94$ & $1.753$ & $.17 - .83$ & $1.876$ & $.61 - .79$ & $1.897$ & \multicolumn{2}{|c|}{$NA$}\\
\hline
$.95$ & $ .04 - .96$ & $1.970$ & $.11 - .89$ & $2.116$ & $.45 - .89$ & $2.165$ & $.60 - .80$ & $2.165$\\
\hline
$1$ & $ .03 - .97$ & $2.057$ & $.08 - .92$ & $2.205$ & $.41 - .92$ & $2.256$ & $.45 - .89$ & $2.279$\\
\hline
\end{tabular}
\caption{[Werner Like States] Complementarity bounds for the change in superdense coding capacity and broadcasting fidelity obtained with a $1-N$ nonlocal cloner for different values of the input state parameter($p$)}
\label{wernerlike4}
\end{table}

\begin{table}[h]
\begin{tabular}{| p{0.8cm} | p{0.8cm} | p{1.3cm} | p{2.5cm} | p{1.3cm} | p{2.5cm}| p{1.3cm} | p{2.5cm}| p{1.3cm} | p{2.5cm} |}
\hline
\multicolumn{2}{| c |}{ No. of copies} & \multicolumn{2}{| c |}{2 copies} & \multicolumn{2}{| c |}{3 copies} & \multicolumn{2}{| c |}{4 copies} & \multicolumn{2}{| c |}{5 copies} \\
\hline
 $c3$  & $c2$ & $c1$ Range & $(\Delta TF +FB)$max & $c1$ Range & $(\Delta TF + FB)$max  & $c1$ Range & $(\Delta TF + FB)$max  & $c1$ Range & $(\Delta TF + FB)$max  \\
\hline
$.56$ & $-.56$ & $.55-1$ & $1.084$ & \multicolumn{2}{| c |}{$NA$} & \multicolumn{2}{| c |}{$NA$} &\multicolumn{2}{| c |}{$NA$} \\
\hline
$.74$ & $-.74$ & $.48-1$ & $1.089$ & $.67-1$ & $1.099$ & \multicolumn{2}{| c |}{$NA$}  &\multicolumn{2}{| c |}{$NA$}  \\
\hline
$.87$ & $-.87$ & $.74-1$ & $1.074$ & $.74-1$ & $1.073$ & $1$ & $.925$ &\multicolumn{2}{| c |}{$NA$}  \\
\hline
$1$ & $-1$ & $1$ & $.9$ & $1$ & $.867$ & $1$ & $.85$ & $1$ & $.84$\\
\hline
\end{tabular}
\caption{[Bell Diagonal States] Complementarity bound for the change in maximal teleportation fidelity and broadcasting fidelity obtained with a $1-N$ nonlocal cloner for different values of the input state parameter($c_2,c_3$)}
\label{belldiagonal1}
\end{table}

\begin{table}[h]
\begin{tabular}{| p{0.8cm} | p{0.8cm} | p{1.3cm} | p{2.5cm} | p{1.3cm} | p{2.5cm}| p{1.3cm} | p{2.5cm}| p{1.3cm} | p{2.5cm} |}
\hline
\multicolumn{2}{| c |}{ No. of copies} & \multicolumn{2}{| c |}{2 copies} & \multicolumn{2}{| c |}{3 copies} & \multicolumn{2}{| c |}{4 copies} & \multicolumn{2}{| c |}{5 copies} \\
\hline
  $c3$  & $c2$ & $c1$ Range & $(\Delta DC +FB)$max & $c1$ Range & $(\Delta DC + FB)$max  & $c1$ Range & $(\Delta DC + FB)$max  & $c1$ Range & $(\Delta DC + FB)$max  \\
\hline
$0.56$ & $-0.56$ & $0.55-1$ & $1.667$ & \multicolumn{2}{| c |}{$NA$} & \multicolumn{2}{| c |}{$NA$} &\multicolumn{2}{| c |}{$NA$} \\
\hline
$0.74$ & $-0.74$ & $0.48-1$ & $1.505$ & $0.67-1$ & $1.888$ & \multicolumn{2}{| c |}{$NA$}  &\multicolumn{2}{| c |}{$NA$}  \\
\hline
$0.87$ & $-0.87$ & $0.74-1$ & $1.900$ & $0.74-1$ & $2.031$ & $1$ & $NA$ &\multicolumn{2}{| c |}{$NA$}  \\
\hline
$1$ & $-1$ & $1$ & $NA$ & $1$ & $NA$ & $1$ & $NA$ & $1$ & $NA$\\
\hline
\end{tabular}
\caption{[Bell Diagonal States] Complementarity bounds for the change in superdense coding capacity and broadcasting fidelity obtained with a $1-N$ nonlocal cloner for different values of the input state parameter($c_2,c_3$)}
\end{table}
\label{belldiagonal2}
\end{widetext}

\end{document}